# Sound transmission loss through double glazing windows in low frequency range

Walid Larbi[1], Chaima Soussi[1], Jean-François Deü[1] and Rubens Sampaio[2]

[1] Conservatoire National des Arts et Métiers (Cnam), Laboratoire de Mécanique des Structures et des Systèmes Couplés (LMSSC), 2 rue Conté, 75003 Paris, France

[2] PUC-Rio, Mechanical Engineering Department, Rua Marques de Sao Vicente, 225 Gavea, Rio de Janeiro, RJ CEP 22453-900, Brazil

*Abstract: The domestic windows in the exterior building facade play a significant role in sound insulation against outdoor airborne noise. The prediction of their acoustic performances is classically carried out in laboratory according to standard ISO 10140. In this work, a 3D elasto-acoustic finite element model (FEM) is proposed to predict the sound reduction index of three different glazing configurations of domestic window, which are compared to laboratory measurements. Two acoustic cavities with rigid-boundaries on both sides of the window are used to simulate respectively the diffuse field on the source side and the pressure field on the receiver side. By using a simplified FEM for the double glazed windows, the sound reduction index is calculated from the difference between the source and receiving sound pressure levels in the one-third octave band from 100 to 500 Hz.*

## INTRODUCTION

The windows in the exterior building facade play an important role in sound insulation against outdoor noise sources such as road traffic and aircrafts in the low frequency range. Therefore, the acoustic performances of domestic window have been the subject of numerous studies to ensure compliance with their sound insulation capabilities.

The principal acoustic indicator is the sound reduction index values *R* measured according to the standards in laboratory practice which takes into consideration two reverberation chambers: a source chamber and a receiving chamber separated by a common wall containing an opening in which the test element is mounted. An important topic in sound insulation measurements is the reproducibility since significant differences have been observed in results from different laboratories especially in the low frequency range (Cops et al. 1987), (Kihlman and Nilsson 1972). For this issue, the niche effect due to the specific positioning of the test window in the wall has been investigated numerically by (Sakuma et al. 2011) and validated with experimental results by (Dijckmans and Vermeir 2012). Results showed that the position of the test specimen in the opening and the depth of the niche have significant influences on the sound transmission loss. Also, research has mainly been focused on evaluation the sound insulation effect of geometry parameters of the window such as type of glass and type of connections between window frame and the wall. (Miskinis et al. 2015) compared experimentally different double glass models and results showed that the best choice is with the combination of one ordinary and one laminated glass. For the same objective, a combined experimental and analytical approach was developed to predict the sound transmission loss of homogenous and laminated glazing by (Ruggeri et al. 2015). In a similar context, a finite element model of a sandwich plate with viscoelastic core was developed by (Larbi et al. 2016) to evaluate the effects of such material on the sound transmission through the studied system. Recently, (Løvholt et al. 2017) studied the effect of window connections with a 3D numerical modeling in the very low frequencies (below 100 Hz). Close agreements with results from laboratory measurements was obtained and showed that the connections have a large importance in the structural vibration and sound transmission. They showed also that the low frequency transmission from 15 to 30 Hz is controlled by the windows, whereas the walls controlled the transmission from 30 to 100 Hz.

This paper is organized as follows: second section describes (i) the test method used for the laboratory measurements required by ISO 10410 and (ii) the acoustic indicators for the evaluation of the sound insulation performances of building elements. The third section is dedicated to the laboratory conditions in which measurements were carried out and to the 3D numerical model. Finally, comparison and discussion between numerical and experimental results of three double glazing windows are proposed.

## EXPERIMENTAL MEASUREMENTS OF SOUND REDUCTION

The airborne sound insulation of building elements such as walls, doors, windows can be evaluated in laboratory according to standards norms such as (ISO 10140, 2010). These standards allow the determination of the acoustic performances of the structures by the measurement of the sound reduction index $R$ (called also Sound Transmission Loss STL in the English-speaking counties).

### Summary of laboratory measurement method

According to different subparts of the standard ISO 10140, the laboratory test measurement takes into consideration two reverberation rooms, horizontally or vertically adjacent, separated by a common wall containing an opening in which the test element is mounted as shown in Fig.1. The minimum volume of two chambers is 50 m³ with a volume difference of 10%. The rooms are mounted on elastic supports and the only significant sound transmission should be through the test element. The reverberant time should not be greater than 2 seconds and the total absorption in the receiving chamber should be low to ensure the best possible condition of the diffuse field. During all measurements, the average relative humidity should be at least 30 % and temperatures should be in the range $20 \pm 3°C$.

In the source room, the sound signals should be random and have a continuous spectrum within each frequency band. For that, one or more loudspeakers can be used, which preferably should be omnidirectional to excite the sound field in the room as uniformly as possible, so the diffuse condition is approximately satisfied. This enables the acoustic pressure to be characterized by a space and frequency averaging which used to calculate the acoustic indicators in one-third octave bands within the frequency range 100 Hz to 5000 Hz as explained in the next section.

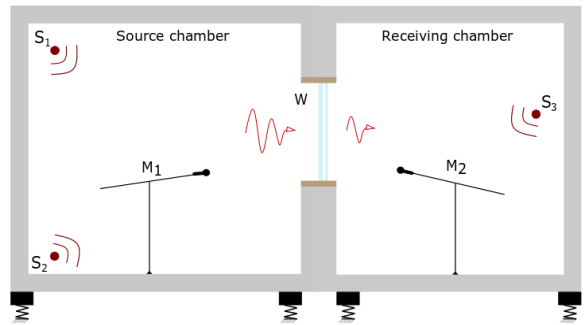

**Figure 1 -** Experimental set-up: Rotating microphones $M_1$ and $M_2$ ; Loudspeakers $S_1$, $S_2$ and $S_3$ ; Tested window W.

### Acoustic indicators

The sound reduction index $R$ is calculated in decibels (dB) as the logarithmic inverse of the sound transmission coefficient $\tau$

$$R = 10 \ \log_{10}\left(\frac{1}{\tau}\right) \quad (1)$$

where $\tau$ is defined as the ratio of the sound power radiated by the test element $w_r$ to the sound power incident $w_i$.

During the laboratory measurements, $R$ is defined as the difference between the average sound pressure levels in the source room and in the receiving room, respectively L$s$ and Lr considering the total absorption of the receiving room $A$ (m²) and the area of the test specimen $S$ (m²):

$$R = L_s\text{-}L_r + 10 \log_{10} \frac{S}{A} \quad (2)$$

The total absorption *A* is determined with Sabine's formula in which the reverberation time $T_r$ (s) in the receiving room and its volume $V$ (m³) are used:

$$T_r = \frac{0.16\, V}{A} \tag{3}$$

$L_s$ and $L_r$ expressed in dB are ten times the logarithm of the ratio of the square of the sound pressure *p* (Pa) in the considering chamber to the square of a reference value, $p_0$ (Pa):

$$L_s = 10 \log_{10} \frac{p^2}{p_0^2} \tag{4}$$

where the reference value $p_0$ is 20 µPa.

**EXPERIMENTAL AND NUMERICAL EXAMPLES**

*Experimental set-up*

Eight types of double glazing windows have been investigated experimentally, but only the results of three models 6/18/4, 4/20/4 and 10/14/4 are presented in this paper and compared to numerical results. We recall that double glazing units are made up of two pieces of glass with a spacer (i.e. 6/18/4 means that the double glazing unit consists of a 6 mm pane, a 18 mm spacer and a 4 mm pane).

The measurements of the sound insulation performance of windows were done in laboratory considering one-third-octave bands from 100 Hz up to 5000 Hz. The test rooms are composed of a source chamber (volume 73 m³) and a receiving chamber (volume 65 m³). Three loudspeakers (RCF-C5215-W), two rotating microphones (Brüel & Kjær Type 3923) and a multi-channel measuring and analyzing device are used (see Fig. 1). The accuracy of the equipment is verified in an accredited metrology described in the standard ISO 10140-5.

The loudspeakers $S_1$ placed in the top left corner and $S_2$ in the bottom left corner of the source room generate a pink noise with a total power of 100 dB. A combination of two heights of the loudspeaker $S_3$ (1.67 m and 2.045 m from the floor) and three positions of the microphone $M_2$ (120°, 240° and 360°) is used to measure the average of the reverberation time $T_r$.

During the measurements, the test window is fixed with wooden sticks and its perimeter, between the wood frame and the opening, is sealed with silicon in both sides to prevent acoustic leakages. The temperature, pressure and relative humidity are kept constant (respectively 20°C, 1036 Pa and 46%) and controlled using a sensor.

The frame of the window tested (Fig. 2) is 1.45 m wide by 1.48 m high and the frame thickness is 47mm. The frame material is Sapele wood with density of 690 kg/m³ and elastic modulus of 14 GPa. The glass material has density of 2500 kg/m³.

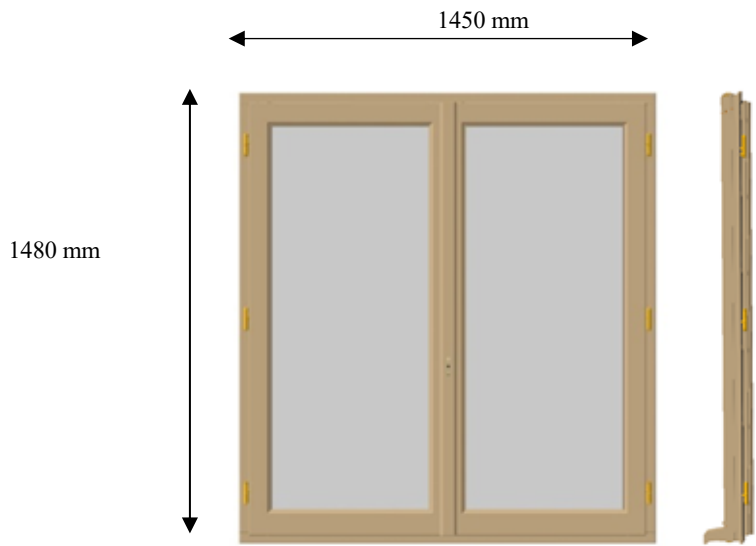

**Figure 2 - The CAD model of test window**

*Numerical model*

We describe in this section the numerical model used for the simulations. On usual assumptions, a vibro-acoustic problem couples a structure (elastic material) to an acoustic domain (fluid). In the finite element context, usually such a model is described in terms of structure displacement $u$ and acoustic pressure $p$. The discretization of the weak formulation of the problem in the frequency domain leads to the following system of equations:

$$\left( \begin{bmatrix} K_s & -C \\ 0 & K_a \end{bmatrix} + i\omega \begin{bmatrix} D_s & 0 \\ 0 & D_a \end{bmatrix} - \omega^2 \begin{bmatrix} M_s & 0 \\ C^T & M_a \end{bmatrix} \right) \begin{bmatrix} u(\omega) \\ p(\omega) \end{bmatrix} = \begin{bmatrix} f_s(\omega) \\ f_a(\omega) \end{bmatrix} \quad (5)$$

where $K_s$, $M_s$ and $D_s$ are the structural stiffness, mass and damping matrices, $K_a$, $M_a$ and $D_a$ are the associated acoustic matrices. $C$ is the coupling matrix, $f_s$ and $f_a$ are the structural and acoustic loads, respectively.

In the present study, the vibro-acoustic system is composed of the window and the two rooms containing air. The 3D finite element model is carried out, as shown in Fig. 3, composed of a source and receiving chambers which are modeled as two rigid acoustic cavities and have the same dimensions of the laboratory configuration. At this stage, a simplified configuration of the window frame was used; a double ordinary glazing (1.45 m wide by 1.48 m high) modeled as clamped thin shells separated by an airtight cavity which is filled with argon whose sound speed is 317 m/s and the density is 1.6 kg/m³. A diffuse sound field is modeled as the acoustic excitation in the source room.

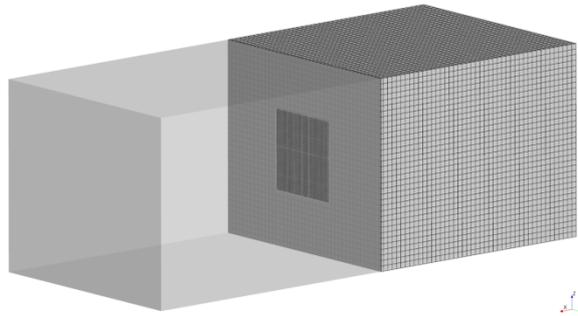

**Figure 3 - Numerical FE model**

A direct frequency analysis is used to compute the response of this model. It covered the range from 0 to 600 Hz with 2 Hz steps and results are presented in the one-third octave band from 100 to 500 Hz. It can be noted that the output noise reduction is given for each frequency and one-third-octave filter was needed to transform results.

Linear finite elements were employed throughout the model. The mesh sizes for the structure and the acoustic domain are controlled by the wavelength $\lambda$ (m) which depends on the frequency range of interest. For the acoustic domain, the acoustic wavelength is $\lambda_a = c/f$ where $c$ (m/s) is the speed of sound. For the structure, the flexural wavelength is $\lambda_f = \sqrt{2\pi/f} \, (D/M)^{1/4}$ where $D$ (Pa.m³) is the flexural rigidity and $M$ (kg/m²) the surface mass density. Since it is recommended to use 6 elements per wavelength, we obtain a model containing around $1.5 \, 10^5$ degrees of freedom.

*Results and discussion*

Numerical third-octave band results are compared to experimental results within the frequency range 100 Hz to 500 Hz (8 frequency values). For the three systems studied 10/14/4, 4/20/4 and 6/18/4, we may conclude from Fig. 4 that, overall, experimental and numerical results follow similar pattern but there are some differences especially in the low frequency range. It is important to keep in mind that experimental data are for the global structure of windows while the numerical results are for simplified double glazing plates. In fact, the overall sound insulation of the window is affected by the frame, its various materials, the sealing of the glazing and other parameters. In addition, the reverberant time $T_r$ measured during the laboratory tests is around 1.5 s which leads to a crossover frequency "Schroeder frequency" $f_s$ about 280 Hz. According to (Schroeder 1996), for airborne sound in reverberant room with a volume $V$, this frequency is given by $f_s = 2000\sqrt{T_r/V}$, and it marks the limit between the well-separated resonances and many overlapping normal modes (diffuse mode). So that, the difference between experimental and numerical results in the range frequency below 300 Hz can be explained by the fact that the sound field is not perfectly diffused.

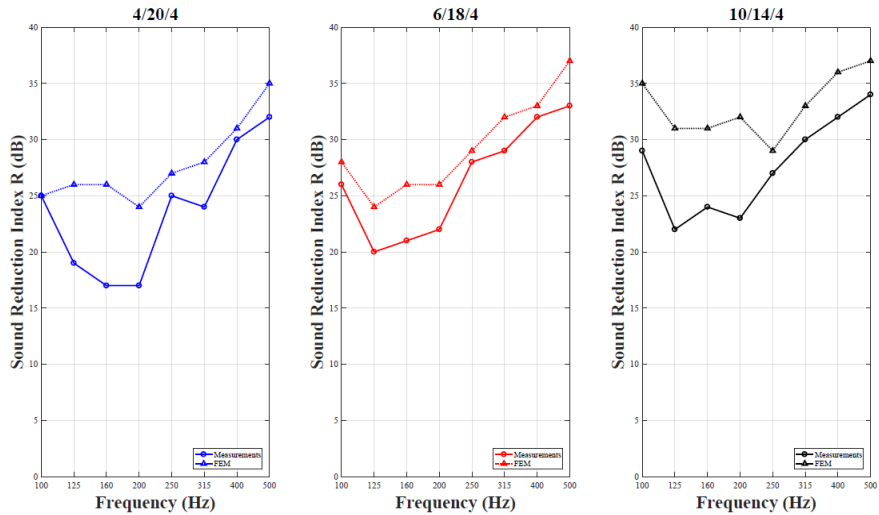

**Figure 4 - Numerical and experimental results for the three types of window**

Fig. 5 shows the effect of the type of glazing systems in the acoustic insulation performance. This figure presents experimental and numerical sound reduction index $R$ for the three glazing studied in this project. The sound transmission loss of a double plate with a fluid gap is classically characterized by four domains. The first one is called the stiffness controlled region which is located at the range of frequencies below the first resonant frequency of the structure. In this domain, the system with higher mass per unit area has the higher $R$. This behavior is confirmed in Fig. 5. The 10/14/4 which is the heaviest has the best acoustic performances and the 4/20/4 has the lowest $R$. The second region is the mass-air-mass resonant frequency. In this region, the sound reduction index is reduced due to the coupling of the two plates with the fluid gap. Differences between the experimental and numerical results can be due to the boundary conditions which have an important effect in the resonant frequency of the system. In fact, in the present numerical model the two plates are clamped on their edges, which is not perfectly representative of real boundary conditions.

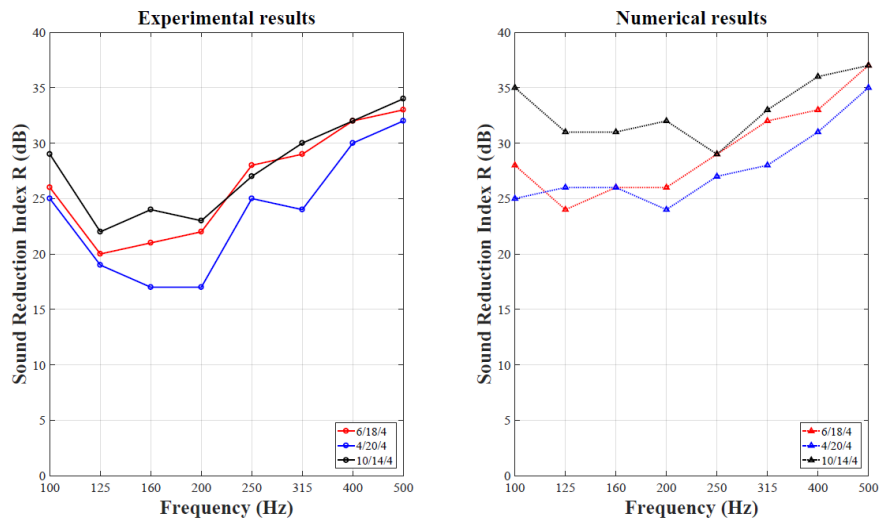

**Figure 5 - Effect of the type of glazing systems in the sound reduction index**

## CONCLUSION

The main objective of this work is to propose an efficient numerical tool based on the finite element method to evaluate the acoustic performances of windows in the exterior building facade. In a first approach, the window is modeled as two glazing plates with an argon gap. Finite elements results for three glazing configurations are obtained and compare to laboratory measurements showing a relatively good agreement between numerical and experimental sound reduction index. Further investigations will concern introduction of passive dissipation in the fluid and active one in the structure using piezoelectric materials (Larbi et al. 2014).